\documentclass[aps,pre,twocolumn,nofootinbib,showkeys]{revtex4}
\usepackage{natbib}
\usepackage[english]{babel}
\usepackage[utf8]{inputenc}
\pdfoutput=1
\usepackage{amsmath}
\usepackage{amssymb}

\usepackage{graphicx}
\usepackage{float}
\usepackage{multirow}

\usepackage{color}

\begin{document}

\author{Julien Varennes}
\affiliation{Department of Physics and Astronomy, Purdue University, West Lafayette, IN 47907, USA}

\author{Andrew Mugler}
\email{amugler@purdue.edu}
\affiliation{Department of Physics and Astronomy, Purdue University, West Lafayette, IN 47907, USA}

\title{ Sense and sensitivity: physical limits to multicellular sensing, migration and drug response}

\begin{abstract}
    Metastasis is a process of cell migration that can be collective and guided by chemical cues. Viewing metastasis in this way, as a physical phenomenon, allows one to draw upon insights from other studies of collective sensing and migration in cell biology. Here we review recent progress in the study of cell sensing and migration as collective phenomena, including in the context of metastatic cells.
    We describe simple physical models that yield the limits to the precision of cell sensing, and we review experimental evidence that cells operate near these limits. Models of collective migration are surveyed in order understand how collective metastatic invasion can occur.
    We conclude by contrasting cells' sensory abilities with their sensitivity to drugs, and suggesting potential alternatives to cell-death-based cancer therapies.
\end{abstract}


\maketitle

\newpage

\begin{figure}[tb]
    \centering
        \includegraphics[width=0.9\columnwidth]{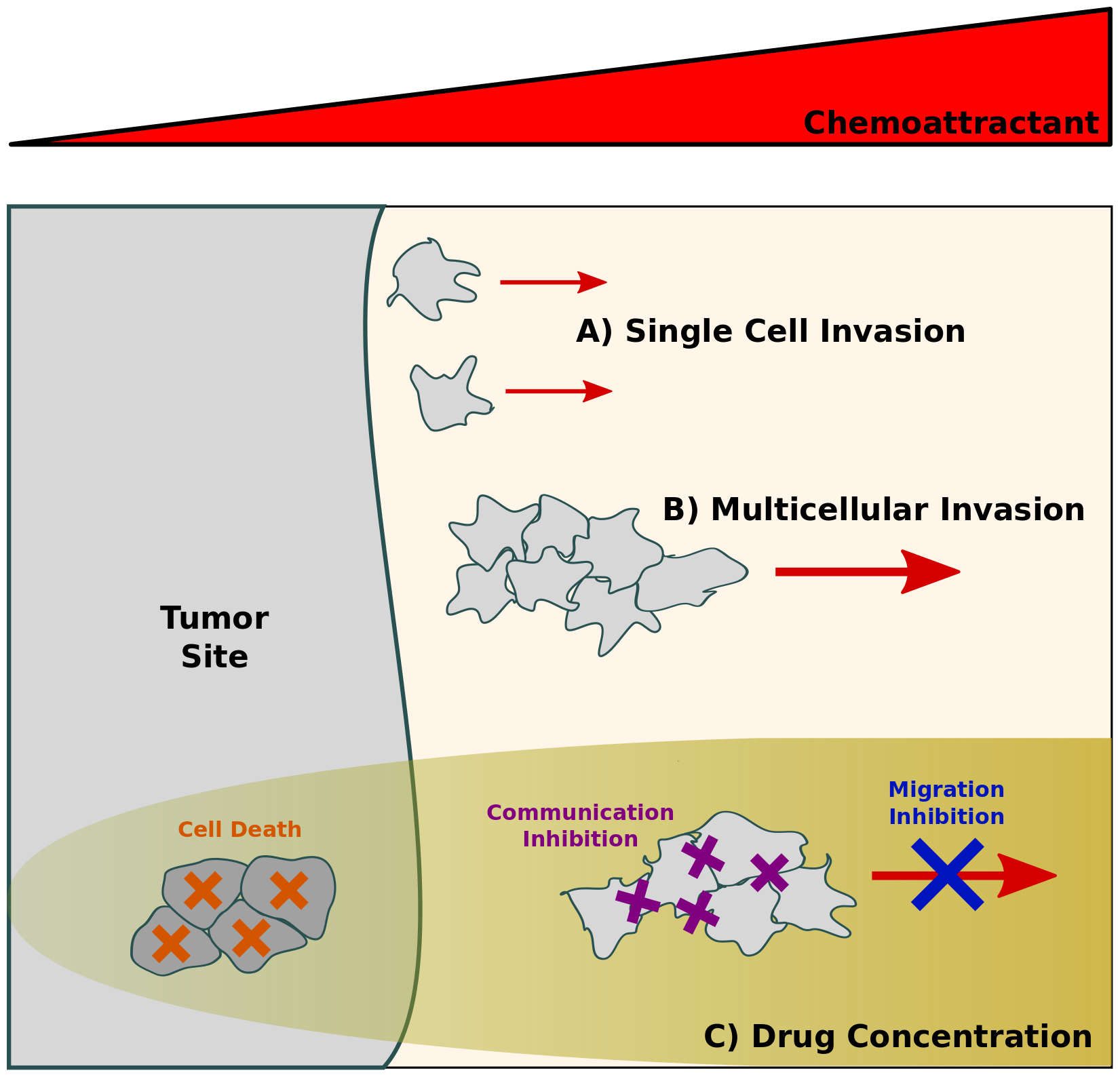}
    \caption{Metastatic invasion is guided by chemical attractants and can occur via (A) single cells or (B) multicellular groups. (C) Drugs are delivered to the tumor environment in order to prevent tumor growth and metastasis. Drugs may cause cell death (orange), block cell-to-cell communication (purple), or prevent cell migration (blue).}
\label{overview}
\end{figure}

Metastasis is one of the most intensely studied stages of cancer progression because it is the most deadly stage of cancer. The first step of metastasis is invasion, wherein cells break away from the tumor and invade the surrounding tissue. Our understanding of metastatic invasion has benefited tremendously from genetic and biochemical approaches \cite{leber2009molecular, hanahan2000hallmarks, hanahan2011hallmarks}. However, the physical aspects of metastatic invasion are still unclear \cite{hanahan2011hallmarks}. We know that at a fundamental level, metastatic invasion is a physical process. Tumor cells sense and respond to chemical gradients provided by surrounding cells \cite{bhowmick2004stromal, condeelis2006macrophages, shields2007autologous, puliafito2015three} or other features of the tumor environment \cite{shields2007autologous, polacheck2011interstitial, shieh2011regulation} (Fig.\ \ref{overview}A). Indeed, tumor cells are highly sensitive, able to detect a $1\%$ difference in concentration across the cell length \cite{shields2007autologous}. Sensing is ultimately a physical phenomenon. Therefore, can we build a simple physical theory to understand the sensory behavior of tumor cells, and can this physical theory inform treatment options?

Metastatic invasion involves coordinated migration of tumor cells away from the tumor site. In many types of cancer, migration is collective and highly organized, involving the coherent motion of connected groups of cells \cite{cheung2013collective, friedl2012classifying, aceto2014circulating, puliafito2015three} (Fig.\ \ref{overview}B). Collective migration is ultimately a physical phenomenon, since it relies on mechanical coupling and can often be understood as emerging from simple physical interactions at the cell-to-cell level. Can we understand the collective migration of tumor cells with simple physical models?

Here we review recent progress on modeling sensing and migration in cells and cell collectives. We discuss metastatic cells explicitly, and emphasize that physical insights gained from other cellular systems can inform our understanding of metastatic invasion.
We focus on simple physical models and order-of-magnitude numerical estimates in order to quantitatively probe the extent of, and the limits to, cell sensory and migratory behavior.
Our hope is that a more quantitative understanding of metastatic invasion will inform treatment protocols, and to that end we conclude by discussing drug sensitivity and potential treatment strategies (Fig.\ \ref{overview}C).

\section{Physical limits to sensory precision}

Tumor cells sense very small concentration gradients\cite{shields2007autologous} and act in a collective manner\cite{cheung2013collective, friedl2012classifying, aceto2014circulating, puliafito2015three}. Here we review the basic theory of concentration and gradient sensing by cells and cell collectives. This theory places physical bounds on sensory precision and allows us to quantitatively compare the capabilities of tumor cells to other cell types.

\subsection{Single-cell concentration sensing}

\begin{figure*}[tb]
    \centering
        \includegraphics[width=.75\textwidth]{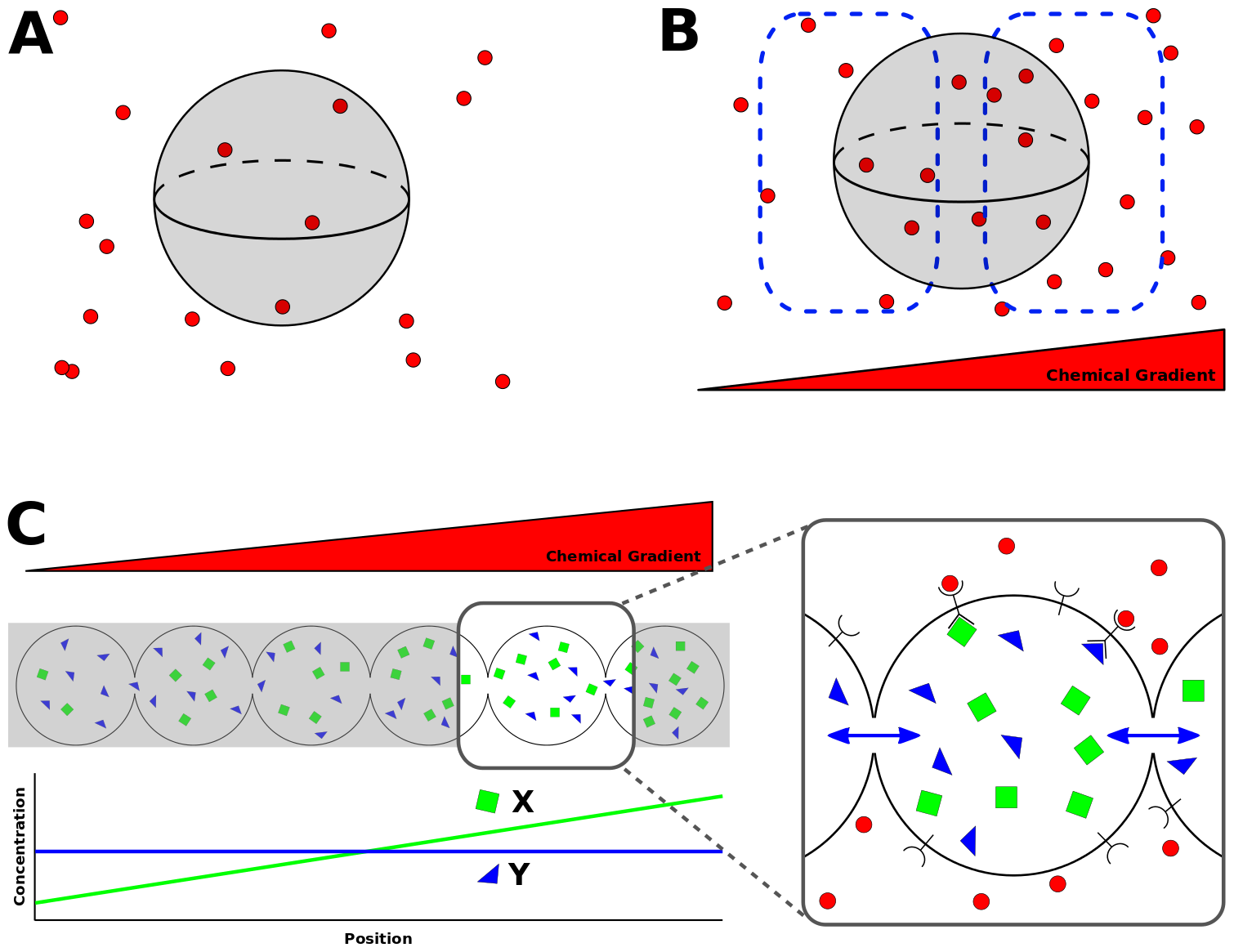}
    \caption{
    (A) An idealized cell as a permeable sphere that counts molecules inside its volume.
    (B) A cell counts molecules in two compartments in order to estimate a concentration gradient.
    (C) The local excitation--global inhibition (LEGI) model of multicellular gradient sensing. $Y$ molecules diffuse between neighboring cells, whereas $X$ molecules do not. The difference between $X$ and $Y$ counts in a given cell reports the extent to which that cell's concentration measurements are above the average.}
    \label{sensing}
\end{figure*}

Theoretical limits to the precision of concentration sensing were first introduced by Berg and Purcell almost 40 years ago\cite{berg1977physics}. Berg and Purcell began by considering an idealized cell that acts as a perfect counting instrument. Their simplest model assumed that the cell is a sphere in which molecules can freely diffuse in and out (Fig.\ \ref{sensing}A). The concentration of these molecules is uniform in space, and the cell derives all its information about the concentration by counting each molecule inside its spherical body. The expected count is
$\bar{n} = \bar{c}V$ where $\bar{c}$ is the mean concentration and $V$ is the cell volume.
However, since molecules arrive and leave via diffusion, there will be fluctuations around this expected value. Diffusion is a Poisson process, meaning that the variance in this count $\sigma_n^2$ equals the mean $\bar{n}$. Therefore the relative error in the cell's concentration estimate is
$\sigma_c/\bar{c} = \sigma_n/\bar{n} = 1/\sqrt{\bar{c} V}$.

The cell can improve upon the relative error in its concentration estimate by time-averaging over multiple measurements. However, consecutive measurements are only statistically independent if they are separated by a sufficient amount of time such that the molecules inside the cell volume are refreshed. The amount of time required is characterized by the diffusion time, $\tau \sim V^{2/3}/D \sim a^2/D$, where $D$ is the diffusion constant and $a$ is the cell diameter. In a time period $T$ the cell makes $\nu = T/\tau$ independent measurements, and the variance is reduced by the factor $1/\nu$. This gives the long-standing lower limit
\begin{equation} \label{eq:singleConc}
\frac{ \sigma_c }{\bar{c}} = \frac{ \sigma_n}{\bar{n}} \sim \frac{1}{\sqrt{a\bar{c}DT}}
\end{equation}
for the cell's relative error in estimating a uniform concentration.
The relative error decreases with $a$ and $\bar{c}$, since the molecule count is larger, and also with $D$ and $T$, since more independent measurements can be made. Berg and Purcell derived this limit more rigorously\cite{berg1977physics}, and the problem has been revisited more recently to account for binding kinetics, spatiotemporal correlations, and spatial confinement \cite{bialek2005physical, kaizu2014berg, bicknell2015limits}. In all cases a term of the form in Eq.\ \ref{eq:singleConc} emerges as the fundamental limit for three-dimensional diffusion.

Do cells reach this limit? Berg and Purcell themselves asked this question in the context of several single-celled organisms, including the \textit{Escheria coli} bacterium \cite{berg1977physics}. Motility of \textit{E.\ coli} has two distinct phases: the run phase in which a cell swims in a fixed direction, and the tumble phase in which the cell erratically rotates in order to begin a new run in a different direction.
The bacterium biases its motion by continually measuring the chemoattractant concentration, and extending the time of runs for which the change in concentration is positive \cite{celani2010bacterial,tu2013quantitative,berg1977physics,dahlquist1976studies}.
The change in concentration $\Delta \bar{c} = Tv\bar{g}$ over a run time $T$ depends on the concentration gradient $\bar{g} = \partial \bar{c} / \partial x$ and the bacterium's velocity $v$.
Berg and Purcell argued that for a change in concentration to be detectable, it must be larger than the measurement uncertainty,
$\Delta \bar{c} > \sigma_c $. Together with Eq.\ \ref{eq:singleConc}, this places a lower limit on the run time,
$T > [ \bar{c}/(aDv^2\bar{g}^2)]^{1/3}$.
Using typical values \cite{berg1977physics} for the sensory threshold of \textit{E.\ coli} of $\bar{c} = 1$ mM, $\partial \bar{c} / \partial x = 1$ mM/cm, $a=1$ $\mu$m, $v=15$ $\mu$m/s, and $D=10^{-5}$ cm$^2$/s, we find $T > 0.1$ s. Actual run times are on the order of $1$ s. Thus we see that \textit{E.\ coli} chemotaxis is consistent with this physical bound.
Although the end goal of concentration sensing in \textit{E.\ coli} is chemotaxis by temporally sampling changes in the chemical concentration, we would like to focus the reader's attention on the remarkable fact that the bacterium's concentration sensing machinery operates very near the predicted physical limits. If \textit{E.\ coli} were to use any shorter run times, chemotaxis would be physically impossible. Consequently, the time period for measuring the chemical concentration, $T$ in Eq.\ \ref{eq:singleConc}, would be so short that the bacterium would be unable to make an accurate measurement of the chemical concentration.

\subsection{Single-cell gradient sensing}

Cells are not only able to detect chemical concentrations, they are also able to measure spatial concentration gradients. Many cells, including amoeba, epithelial cells, neutrophils, and neurons, sense gradients by comparing concentrations between compartments in different locations of the cell body \cite{jilkine2011comparison}.
These compartments are typically receptors or groups of receptors on the cell surface, but in a simple model we may treat these compartments as idealized counting volumes as we did before (Fig.\ \ref{sensing}B). The difference in counts between two such compartments provides the cell with an estimate of the gradient. What is the relative error in this estimate?

Consider two compartments of linear size $s$ on either side of a cell with diameter $a$ (Fig.\ \ref{sensing}B). If the compartments are aligned with the gradient $\bar{g}$ of a linear concentration profile, then the mean concentrations at each compartment are $\bar{c}_1$ and
$\bar{c}_2 = \bar{c}_1 + a\bar{g}$.
The mean molecule counts in the two compartments are roughly $\bar{n}_1 = \bar{c}_1s^3$ and $\bar{n}_2 = \bar{c}_2s^3$, and the difference is $\Delta\bar{n} = \bar{n}_2 - \bar{n}_1 = a\bar{g}s^3$. The variance in this difference is
$\sigma_{\Delta n}^2 = \sigma_{n_1}^2 + \sigma_{n_2}^2 \sim \bar{n}_1^2/(s\bar{c}_1DT) + \bar{n}_2^2/(s\bar{c}_2DT)$,
where the first step assumes the two compartments are independent, and the second step uses Eq.\ \ref{eq:singleConc} for the variance in each compartment's measurement. For shallow gradients, where the limits on sensing are generally probed, we have $a\bar{g} \ll \bar{c}_1$, and therefore we may assume $\bar{c_1} \approx \bar{c_2} \approx \bar{c}$, where $\bar{c}$ is the mean concentration at the center of the cell. Thus
$\sigma_{\Delta n}^2 \sim 2(\bar{c}s^3)^2/(s\bar{c}DT)$,
and the relative error in the cell's estimate of the gradient is then
\begin{equation}
\label{eq:g}
\frac{\sigma_g}{\bar{g}} = \frac{\sigma_{\Delta n}}{\Delta \bar{n}} \sim \sqrt{\frac{\bar{c}}{s(a\bar{g})^2DT}},
\end{equation}
where the factor of $2$ is neglected in this simple scaling estimate. As in Eq.\ \ref{eq:singleConc}, we see that the relative error decreases with $s$, since the molecule counts in each compartment are larger, and also with $D$ and $T$, since more independent measurements can be made. Additionally, the relative error decreases with $a\bar{g}$, since the concentrations measured by the two compartments are more different from each other. However, we see that unlike in Eq.\ \ref{eq:singleConc}, the relative error increases with the background concentration $\bar{c}$. The reason is that the cell is not measuring a concentration, but rather a \textit{difference} in concentrations, and it is more difficult to measure a small difference on a larger background than on a smaller background \cite{ellison2016cell}. Eq.\ \ref{eq:g} has been derived more rigorously\cite{endres2009accuracy}, and the problem has been extended to describe rings of receptors \cite{endres2009accuracy} or detectors distributed over the surface of a circle \cite{hu2010physical} or a sphere \cite{endres2008accuracy}. In all cases a term of the form in Eq.\ \ref{eq:g} emerges as the fundamental limit, with the lengthscale $s$ dictated by the particular sensory mechanism and geometry. It is clear that the optimal mechanism would result in an effective compartment size that is roughly half of the cell volume, in which case $s\sim a$.

Do cells reach this limit on gradient sensing? This question has been directly addressed for the amoeba \textit{Dictyostelium discoideum}. Experiments\cite{van2007biased} have shown that \textit{Dictyostelium} cells exhibit biased movement when exposed to gradients of cyclic adenosine monophosphate as small as $\bar{g} = 10$ nM/mm, on top of a background concentration of $\bar{c} = 7$ nM. Bias is typically quantified in terms of the chemotactic index (CI), which is the cosine of the angle between the gradient direction and the direction of a cell's actual motion. By relating the error in gradient sensing (a term of the form in Eq.\ \ref{eq:g} with $s = a$) to the error in this angle, Endres and Wingreen \cite{endres2008accuracy} obtained an expression for the optimal CI, which they then fit to the experimental data with one free parameter, the integration time $T$. The inferred value of $T = 3.2$ s serves as the physical lower bound on the response time required to perform chemotaxis. Actual response times of \textit{Dictyostelium} cells, as measured by the time from the addition of a chemoattractant to the peak activity of an observable signaling pathway associated with cell motility \cite{postma2003uniform, parent2004making}, are about $5$$-$$10$ s. Taken together, these results imply that \textit{Dictyostelium} operates remarkably close to the physical limit to sensory precision set by the physics of molecule counting.

\subsection{Relative changes vs.\ absolute molecule numbers}

The precision of gradient sensing is often reported in terms of percent concentration change across a cell body. For example, both amoeba \cite{van2007biased} and tumor cells \cite{shields2007autologous} are sensitive to a roughly $1\%$ change in concentration across the cell body. However, this method of reporting sensitivity may be misleading. Experiments imply very different sensory thresholds for these cells in terms of absolute molecule numbers, as we will now see.

The key is that it takes two numbers to specify the conditions for gradient sensing: the mean gradient $\bar{g}$ and the mean background concentration $\bar{c}$. For the amoeba \textit{Dictyostelium}, these numbers are $\bar{g} = 10$ nM/mm and $\bar{c} = 7$ nM at the sensory threshold \cite{van2007biased}. Given a typical cell size of $a = 10$ $\mu$m, these values imply a mean percent concentration change of $\bar{p} = a\bar{g}/\bar{c} = 1.4\%$ (Table \ref{sense_table}). However, we may also compute from these values the mean molecule number difference
$\Delta\bar{n} = a\bar{g}s^3$
from one side of the cell to the other, within the effective compartments of size $s$. Taking $s\sim a$ gives the maximal molecule number difference of $\Delta\bar{n} = a^4\bar{g} = 60$
for \textit{Dictyostelium} (Table \ref{sense_table}). Together $\bar{p}$ and $\Delta\bar{n}$ specify the sensing conditions as completely as $\bar{g}$ and $\bar{c}$ do.

Experiments \cite{shields2007autologous} have shown that breast cancer tumor cells exhibit a chemotactic response in a gradient $\bar{g} = 550$ nM/mm of the cytokine CCL21, on top of a background concentration of $\bar{c} = 1100$ nM. Given a typical cell size of $a = 20$ $\mu$m, this corresponds to a percent difference of $\bar{p} = a\bar{g}/\bar{c} = 1\%$, similar to \textit{Dictyostelium}. Yet, this also corresponds to a maximal molecule number difference of $\Delta\bar{n} = a^4\bar{g} = 53,$$000$, which is much higher than that of \textit{Dictyostelium} (Table \ref{sense_table}). Even though the sensitivities are similar in terms of percent change, they are very different in terms of absolute molecule number.

Lower molecule numbers correspond to higher relative error. We can see this explicitly by writing Eq.\ \ref{eq:g} in terms of the percent change $\bar{p} = a\bar{g}/\bar{c}$. Defining $\epsilon = \sigma_g/\bar{g}$ and taking $s \sim a$, we have $\epsilon \sim 1/\sqrt{\bar{p}^2a\bar{c}DT}$. Accounting for the fact that tumor cells (TC) have roughly twice the diameter as \textit{Dictyostelium} cells (DC), this expression implies that the sensitivities of the two cell types over the same integration time $T$ to chemoattractants with the same diffusion constant $D$ satisfy
$\epsilon_{\rm DC}/\epsilon_{\rm TC} = \sqrt{2\bar{c}_{\rm TC}/\bar{c}_{\rm DC}} \approx 18$.
We see that because the \textit{Dictyostelium} experiments were performed at lower background concentration, corresponding to lower absolute molecule numbers, the relative error in gradient sensing is $18$ times that of the tumor cells, despite the fact that both cell types are responsive to $1\%$ concentration gradients. Therefore, it is important to take note of the background concentration when studying the precision of gradient sensing. These data imply that \textit{Dictyostelium} cells can sense noisier gradients than tumor cells. However, \textit{Dictyostelium} cells have been studied more extensively than tumor cells as exemplars of gradient detection. It remains an interesting open question what is the minimum gradient that tumor cells can detect, not only in terms of percent concentration change, but also in terms of absolute molecule number differences.

We see that although cancerous cells and \textit{Dictyostelium} cells are of similar size, their sensory responses to absolute molecule numbers can be very different (Table \ref{sense_table}). This difference is also reflected in their migration speeds: carcinoma and epithelial cells migrate \cite{wolf2003compensation,wang2004differential,gilles1999vimentin,legrand1999airway} at $\sim 0.5 \mu\text{m}/\text{s}$ whereas \textit{Dictyostelium} can migrate \cite{mccann2010cell,song2006dictyostelium} at speeds of $\sim 10 \mu\text{m}/\text{s}$.

\begin{table*}[t]
\begin{center}
\begin{tabular}{ |l|c|c|c|c| }
\hline
\multirow{2}{*}{ }
& \multicolumn{2}{ |c| }{Single Cell} & \multicolumn{2}{ |c| }{Multicellular} \\ \cline{2-5}
 & \textit{Dictyostelium}
 & Breast
 & Neurons \cite{rosoff2004new}
 & Mammary \\
& (Amoeba) \cite{van2007biased}
& Cancer \cite{shields2007autologous}
& & Epithelia \cite{ellison2016cell} \\
\hline
Cell Length & 10 $\mu$m & 20 $\mu$m & 10 $\mu$m & 10 $\mu$m \\
Scale, $a$  & & & & \\ \hline
Background & 7 nM & 1100 nM & 1 nM & 2.5 nM \\
Concentration, $\bar{c}$  & & & & \\ \hline
Concentration & 10 nM/mm & 550 nM/mm & $0.1$ nM/mm & $0.5$ nM/mm \\
Gradient, $\bar{g}$ & & & & \\ \hline
Percent Concentration & 1.4\% & 1.0\% & 0.1\% & 0.2\% \\
Difference, $\bar{p} = \bar{g}a/\bar{c}$ &  &  &  &  \\ \hline
Molecule Number & 60 & 53,000 & 0.6 & 3 \\
Difference, $\Delta\bar{n} = \bar{g}a^4$ & & & & \\ \hline
\end{tabular}
\caption{Gradient sensory thresholds for single cells and multicellular collectives. Note that experiments can provide equal percent concentration differences but unequal molecule number differences across a cell body, as seen for amoeba and breast cancer cells. We see that multicellular groups can detect smaller gradients than single cells by all measures.}
\label{sense_table}
\end{center}
\end{table*}

\subsection{Multicellular gradient sensing}

In many cancer types, tumor cells invade the surrounding tissue in a collective manner \cite{cheung2013collective, friedl2012classifying}. Cell collectives can sense shallower gradients than single cells \cite{ellison2016cell,malet2015collective}, both in terms of percent concentration changes and absolute molecule numbers (Table \ref{sense_table}). Indeed, groups of neurons respond to gradients equivalent to a difference of less than one molecule across an individual neuron's growth cone \cite{rosoff2004new}. This raises the possibility that during the invasion process tumor cell collectives benefit from higher sensory precision than single tumor cells.

We can understand immediately from Eq.\ \ref{eq:g} why a multicellular collective would have lower sensory error than a single cell: a collective is larger than a single cell. Therefore, the collective covers a larger portion of the concentration profile, which leads to a larger difference between the concentration measurements on either end, and a lower relative error. In terms of Eq.\ \ref{eq:g}, if we consider that cells on the ends act as the molecule-counting compartments, $s \to a$, and that the entire collective acts as the detector, $a \to Na$, where $N$ is the number of cells in the gradient direction, then we have \cite{mugler2016limits}
\begin{equation}
\label{eq:G1}
\frac{\sigma_g}{\bar{g}} \sim \sqrt{\frac{\bar{c}}{a(Na\bar{g})^2DT}}.
\end{equation}
We see that, as expected, the relative error goes down with the size $Na$ of the multicellular collective.

However, there is a crucial point that is overlooked in formulating Eq.\ \ref{eq:G1}: the larger the group of cells, the more difficult it is for cells on either end to communicate the measurement information. This fact is not accounted for in Eq.\ \ref{eq:G1}. Instead, we see that the relative error decreases with the separation $Na$ between the end cells without bound, which is unrealistic. For a single cell it may be a reasonable approximation to assume that compartments quickly and reliably communicate information across the cell body, but for a multicellular collective, the communication process cannot be overlooked. Importantly, the communication mechanism of multicellular collectives may introduce additional noise into the gradient sensing process. Therefore, it is imperative when considering collective sensing to properly account for the effects of communication.

Recently, the physical limits to collective gradient sensing including communication effects were derived \cite{ellison2016cell, mugler2016limits}. Communication was modeled using a multicellular version of the local excitation--global inhibition (LEGI) paradigm \cite{levchenko2002models}, in which each cell produces a ``local'' and a ``global'' molecular species in response to the chemoattractant, and the global species is exchanged between cells to provide the communication (Fig.\ \ref{sensing}C). The difference between local and global molecule numbers in a given cell provides the readout. A positive difference informs the cell that its detected chemoattractant concentration is above the spatial average among its neighbors, and therefore that the cell is located up the gradient, not down.
In this model, the relative error of gradient sensing was shown \cite{mugler2016limits} to be limited from below by
\begin{equation}
\label{eq:G2}
\frac{\sigma_g}{\bar{g}} \sim \sqrt{\frac{\bar{c}}{a(n_0a\bar{g})^2DT}},
\end{equation}
where $n_0^2$ is the ratio of the cell-to-cell exchange rate to the degradation rate of the global species. Comparing Eq.\ \ref{eq:G1} to Eq.\ \ref{eq:G2}, we see that without communication the error decreases indefinitely with the size $Na$ of the collective, whereas with communication the error is bounded by that of a collective with effective size $n_0a$. Evidently, communication defines an effective number of cells $n_0$ over which information can be reliably conveyed, and a collective that grows beyond this size no longer improves its sensory precision.

These theoretical predictions were tested experimentally in collectives of epithelial cells \cite{ellison2016cell}. Mouse mammary epithelial cells were grown in organotypic culture and subjected to very shallow gradients of epidermal growth factor (Table \ref{sense_table}). It was shown that while single cells did not respond to these gradients, the multicellular collectives did: they exhibited a biased cell-branching response. Importantly, the response of large collectives was no more biased than the response of small collectives, supporting the theory with communication (Eq.\ \ref{eq:G2}) over the theory without communication (Eq.\ \ref{eq:G1}). The effective detector size was inferred to be $n_0 \approx 3.5$ cells, which is consistent with the size of these collectives in their natural context (the ``end buds'' of growing mammary ducts) \cite{lu2008genetic}. Interestingly, when the gap junctions between cells, which mediate the molecular communication, were blocked with each of several drugs, the biased responses were abolished \cite{ellison2016cell}, demonstrating that the collective response was critically dependent on the cell-to-cell communication. Taken together, these results indicate that cell-to-cell communication is a necessary but imperfect enabler of collective gradient sensing. The results also speak to the power of simple physical theory to quantitatively explain collective cellular capabilities. Since epithelial cancers are known to invade collectively \cite{cheung2013collective}, it remains an important open question whether this theory also describes the sensory behavior of tumor cell collectives.

\section{Physical models of collective migration}

Metastatic invasion is a process of cell migration. Collective invasion, in turn, is a process of collective migration. Therefore, it is important to understand not only the collective sensing capabilities of tumor cells, but also the properties of their collective migration---and ideally the relation between the two. From a physical modeling perspective, describing collective cell dynamics is an interesting problem, because often rich and unexpected behavior can emerge from a few simple interaction rules between cells. Even in the absence of sensing, simple models have successfully explained observed collective behaviors such as cell streaming, cell sorting, cell sheet migration, wound healing, and cell aggregation \cite{kabla2012collective,szabo2010collective,basan2013alignment,janulevicius2015short}. Here we focus on the collective dynamics that emerge when sensing plays a key role. In this case, a sensory cue results in polarization of a cell or cell collective via one of a variety of mechanisms \cite{jilkine2011comparison}, and the dynamics are directed, i.e.\ migratory.

\subsection{Mechanisms of collective migration}

\begin{figure*}[tb]
    \centering
        \includegraphics[width=.75\textwidth]{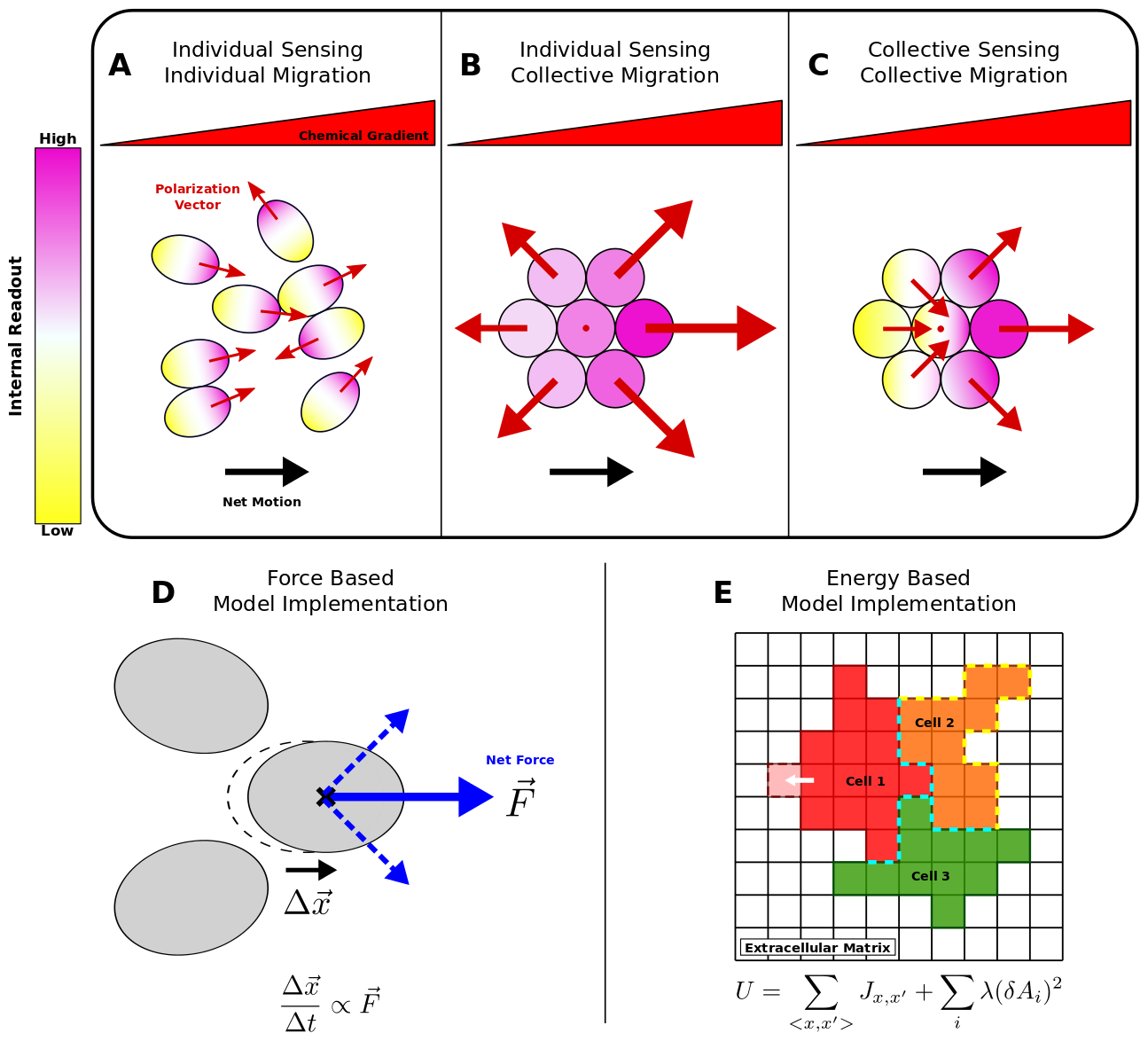}
    \caption{Mechanisms of collective migration: (A) individual sensing and migration (the ``many wrongs'' mechanism), (B) individual sensing but collective migration (emergent chemotaxis), and (C) collective sensing and migration. Implementations of collective migration: (D) in force-based models, dynamics evolve from stochastic forces acting on each cell; (E) in energy-based models, dynamics evolve via energy minimization with thermal noise. E shows the cellular Potts model framework, in which cells are collections of lattice sites, and cell-cell (dashed blue) and cell-environment (dashed yellow) contacts contribute to the energy of the system.}
\label{models}
\end{figure*}

Broadly speaking, the mechanisms of collective migration can be divided into three categories. First, cells may exhibit individual sensing and individual migration (Fig.\ \ref{models}A). Here, each cell can perform gradient sensing and migration individually, although the precision may be low. When many such cells are placed in a group, the group migration can be enhanced and focused by local interactions between the cells. Even if each individual cell has low sensory and migratory precision, the precision of the group as a whole is high due to the interactions.
Collisions act to average over the errors in individual cells' noisy measurements, thereby decoupling group behavior from single-cell properties. This mechanism is often termed ``many wrongs,'' and it is successful at explaining how group migratory behavior emerges from individual agents that act independently \cite{simons2004many,coburn2013tactile}. As discussed later, the failure of a communication-blocking therapy could act as proof that a ``many wrongs'' method of collective migration is at work in tumor cell invasion.

Second, cells may exhibit individual sensing but collective migration (Fig.\ \ref{models}B). In this mechanism, each individual cell senses its own local environment, and tight mechanical interactions result in the emergent directed motion of the entire group. This mechanism is applicable to the collective migration of connected clusters of cells. For example, models of this type were recently developed by Camley et al.\ \cite{camley2015emergent} and by Malet-Engra et al.\ \cite{malet2015collective} to describe behavior seen in clusters of neural crest cells and lymphocytes, respectively. In this model, cells are tightly connected but are polarized away from neighboring cells due to contact inhibition of locomotion (CIL), the physical phenomenon of cells ceasing motion in the direction of cell-cell contact \cite{mayor2010keeping}. Individual cells sense a local chemoattractant concentration and attempt to migrate away from the group with a strength proportional to this concentration. However, the mechanical coupling keeps them together. In the presence of a concentration gradient, the imbalance in their migration strengths results in net directed motion (Fig.\ \ref{models}B). Notably, this mechanism results in directed motion of a cluster even though individual cells cannot execute directed motion alone, since without other cells, there is no CIL to bias the motion.

Third, cells may exhibit collective sensing and collective migration (Fig.\ \ref{models}C). As discussed above, multicellular groups exploit cell-to-cell communication to sense gradients collectively, thereby enhancing the precision of sensing. A feature of this collective sensing, e.g.\ via the multicellular LEGI mechanism discussed above \cite{ellison2016cell, mugler2016limits}, is that each cell has information on the extent to which it is up or down the gradient. Through CIL or other contact-mediated interactions, this information can translate directly into cell polarity, leading to more coherent collective migration than in the previous mechanism (Fig.\ \ref{models}C vs.\ B). In fact, the multicellular LEGI model was used by Camley et al.\ \cite{camley2015emergent} to explore a model of this type. Adding collective sensing to their model of CIL-dependent migration gave the advantage that the repulsive tension on a cell cluster was adaptive and therefore remained constant as the cluster migrated to regions of higher chemical concentration.

\subsection{Model implementations}

To study the above mechanisms quantitatively and compare predictions with experiments, one must turn to mathematical and computational modeling. Models of cell dynamics range from continuum or semi-continuum descriptions, which describe groups of cells as continuous tissues, to individual-based models, which describe cells as individual interacting entities \cite{maclaren2015models}. Physics-driven individual-based models generally fall into two categories: force-based models and energy-based models.

Force-based models (Fig.\ \ref{models}D) typically represent cells as centers of mass or as collections of vertices. Cell dynamics evolve from forces acting on individual cells, which can be stochastic, and arise from internal features such as cell polarity, and external features such as mechanical interactions with other cells \cite{maclaren2015models}. Force-based models are able to reproduce multicellular behavior such as chemotaxis, wound healing, and cell aggregation \cite{camley2015emergent,basan2013alignment,janulevicius2015short}. Parameters are often directly relatable to experimental measurements, and the simplest models are often amenable to exact mathematical analysis \cite{camley2015emergent}.

Energy-based models (Fig.\ \ref{models}E) allow cell dynamics to emerge from the minimization of a potential energy with thermal noise (the so-called Monte Carlo scheme). A widely used example is the cellular Potts model (CPM) \cite{graner1992simulation, swat2012multi}, in which cells are represented as collections of co-aligned ``spins'' on a lattice (Fig.\ \ref{models}E). Cells remain contiguous because it is energetically favorable for neighboring spins to be co-aligned. Biophysical features such as cell shape, cell-cell adhesion, and cell protrusions into the environment are modeled by introducing corresponding terms into the global potential energy. The CPM has successfully reproduced experimental observations of cell sorting, streaming, chemotaxis, and collective migration \cite{kabla2012collective,maree2007cellular,szabo2010collective}. In energy-based models, the parameters are set by calibrating emergent features, such as cell diffusion coefficients or average speeds, with experimental measurements \cite{szabo2010collective}.

Although the physical limits to multicellular sensing are becoming better understood, the physical limits constraining multicellular migration are less clear. This remains an interesting open question, and answering it will require integrating the theories of sensing and communication with the models of collective migration described herein. For tumor cells in particular, an integrated physical theory of sensing and migration would prove immensely useful for identifying the key determinants of invasive capabilities. Identifying these determinants would help pinpoint the ways that these capabilities could be disrupted, using drugs and other therapies, as described next.

\section{Drug sensitivity and implications for therapy}

We have seen that cells, including tumor cells, are remarkably precise sensors of molecules in their environment. This raises the question of how sensitive tumor cells are to drug molecules in their environment. What is the minimum drug concentration required not just for precise detection by a cell, but for causing a phenotypic change, such as cell death?

Experiments have shown that cancer cells are sensitive to very small drug concentrations. For example, lung carcinoma cells were exposed \textit{in vitro} to various concentrations of the anti-cancer drug paclitaxel, also known as taxol, which acts to block mitosis in order to achieve cell death by disrupting microtubule regulation \cite{torres1998mechanisms}. Paclitaxel concentrations as low as $1$ nM were shown to affect microtubule dynamics of the cells. This concentration is commensurate with the smallest background concentrations in which cells can perform gradient sensing (Table \ref{sense_table}).
Assuming a cell length of $20$ $\mu$m, which is typical of carcinoma cells \cite{leber2009molecular}, this concentration corresponds to only a few thousand drug molecules in the volume of a cell (Table \ref{drug_table}).
Evidently lung cancer cells are affected by drug concentrations that are near the fundamental limits of what can be sensed.

\begin{table*}[t]
\begin{center}
    \begin{tabular}{ |l|c|c| }
        \hline
        \multicolumn{1}{|c|}{Outcome} & Drug Concentration & Molecules per Cell \\ \hline
        Physical change \cite{torres1998mechanisms} & $1$ nM & $5$,$000$ \\ \hline
        Cell death \cite{grantab2006penetration} & $10^4$ nM & $5\times 10^7$ \\ \hline
        Cell death, nanoparticle delivery \cite{malam2009liposomes} & $100$ nM & $500$,$000$ \\ \hline
        Communication blockage \cite{ellison2016cell} & $50$ nM & $30$,$000$ \\ \hline
    \end{tabular}
\caption{Drug sensitivity thresholds. Molecules per cell volume are calculated assuming a cubic cell of length $a = 20$ $\mu$m for tumor cells (rows 1-3) and $a = 10$ $\mu$m for epithelial cells (row 4).}
\label{drug_table}
\end{center}
\end{table*}

Although cancer cells may be very sensitive to small drug concentrations, that does not translate to successful treatment. In order to achieve cell death, much larger drug concentrations are required. In the same study on lung carcinoma cells, cell death was observed for drug concentrations on the order of $10$ nM and greater. More typical drug concentrations required for cell death are on the order of micromolars. For instance, it has been shown \textit{in vitro} that anticancer drug concentrations on the order of $10$ $\mu$M are required to kill at least $90\%$ of tumor cells \cite{grantab2006penetration}. With a cell length of $20$ $\mu$m, $10$ $\mu$M corresponds to tens of millions of drug molecules in the volume of a cell, four orders of magnitude greater than drug concentrations required to affect cell functionality (Table \ref{drug_table}). In order to effectively kill a solid tumor, very high drug doses are required.

Complicating matters is the fact that the tumor and its surrounding microenvironment comprise a complex and heterogeneous system. Although most cells in the human body are naturally within a few cell lengths of a blood vessel, due to high proliferation tumor cells may be upwards of tens of cell lengths away from a vessel \cite{minchinton2006drug}. This makes it difficult for drugs to reach the tumor. Moreover, the high density of many solid tumors causes gradients of drug concentration to form as a function of tumor radius \cite{kwon2012analysis}. This results in a reduced drug concentration at the center of the tumor and makes innermost tumor cells the most difficult to kill. A promising way to overcome this difficulty is through the use of nanoparticle drug delivery systems, which increase both the specificity and penetration of drugs to the tumor. Nanoparticle delivery has been shown \cite{malam2009liposomes} to achieve cell death with concentrations as low as $100$ nM. Although this concentration is lower than delivery without nanoparticles, it is still two orders of magnitude higher than the minimum concentration that causes physical changes in the cell (Table \ref{drug_table}). Even with targeted delivery, achieving drug-induced tumor cell death remains a challenging task.

Given this challenge, we hope to draw upon the physical insights reviewed herein to devise therapeutic strategies that are alternative or complementary to comprehensive cell death. Specifically, we imagine focusing on the metastatic invasion phase, and targeting the functions of invading tumor cells, including communication and migration, in addition to targeting cells' overall viability, to produce better treatment (Fig.\ \ref{overview}C). Communication is a particularly promising candidate, since it has recently been shown that cell-to-cell communication makes cancer cells more resistant to therapy and helps sustain tumor growth \cite{boelens2014exosome}. Indeed, the exchange of extracellular vesicles, which is a form of communication observed between tumor cells and stromal cells, has been linked to immune suppression, tumor development, angiogensis, and metastasis \cite{vader2014extracellular}. This suggests that disrupting cell-to-cell communication could be an effective strategy for stopping tumor progression or curbing metastatic invasion. Disrupting communication may not require concentrations as large as those necessary for cell death, which are difficult to maintain \textit{in vivo} across the whole tumor. For example, as little as $50$ nM of the gap-junction-blocking drug Endothelin-1 is sufficient to remove collective responses in epithelial cells \cite{ellison2016cell}. This concentration is several orders of magnitude smaller than that required for comprehensive cell death, and it is on the order of concentrations that are effective with targeted nanoparticle delivery (Table \ref{drug_table}). Therefore, it is tempting to suggest that managing metastatic invasion by blocking communication or other cell functions is a more accessible therapeutic strategy than eradicating a tumor outright.

Although blocking intercellular communication pathways could curb the invasive potential of metastatic cells it is also important to address the ulterior consequences of this strategy. Gap junction intercellular communication (GJIC) is an important way for the environment to affect change on cells, maintaining tissue homeostasis and balancing cell growth and death \cite{krysko2005gap}. In cancerous cells GJIC is reduced, causing unregulated cell growth \cite{salameh2005pharmacology}. Interestingly, many existing cancer-combatting drugs are small enough to pass through cell gap junctions which permit molecules of sizes up to 1000 Dalton, but there is a lack of \textit{in vivo} studies concerning the benefits and effects of gap junctions on cancer treatment \cite{salameh2005pharmacology}. It has been shown \textit{in vitro} that GJIC can propagate cell-death signals through cancerous cells and that high connexin expression, the proteins that compose gap junctions, corresponds to high anticancer drug sensitivity \cite{krutovskikh2002gap,krysko2005gap}. Therefore, it is important to consider the potential negative consequences of blocking intercellular communication in reducing metastatic invasion. It may be sufficient to administer an anticancer drug and a communication-blocking drug at different times in order to avoid negative side-effects. Although this puts a caveat on our proposal of communication-blocking drugs as a viable option for treating metastatic invasion it is important to recall that GJIC is not the only communication pathway available to cancerous cells: extracellular vesicle-meditated signaling pathways are potential alternates which could be targeted in place of GJIC \cite{boelens2014exosome,vader2014extracellular}.

\section{Outlook}

In this review, we have taken a quantitative look at metastatic invasion as a sensing-and-migration process, which has allowed us to compare metastatic cells to other cell types in terms of their physical capabilities. We have seen that tumor cells can sense very shallow chemoattractant gradients, which may help guide metastatic invasion, but it remains unclear whether tumor cells  operate near fundamental sensing limits, as bacteria and amoeba do. Recognizing that metastatic invasion can be collective, we have reviewed recent results on the physical limits to collective sensing, and we have identified the overarching mechanisms of collective migration. A key insight that emerges is that collective capabilities rely critically on cell-to-cell communication. This insight opens up alternative strategies for therapy that target specific cell capabilities such as communication, in addition to strategies that aim for comprehensive cell death.

A detailed presentation of the underlying physical mechanics for cell motility and chemotaxis are outside the scope of this review. Readers interested in these topics are referred to these excellent resources \cite{lauffenburger1996cell,van2004chemotaxis,firtel2000molecular,chung2001signaling}. It is also important to note that in deriving the limits to concentration sensing we have assumed that the molecules of interest diffuse normally with fixed, space-independent diffusion coefficients. However, this may not always be the case in the tumor environment, where molecules can also experience anomalous diffusion \cite{chauhan2011delivery,dix2008crowding}.

Moving forward, it will be important to identify whether the physical theory of sensing reviewed herein can be applied in a predictive manner to tumor cells, and whether gradient sensing plays a dominant role during metastatic invasion. More generally, it will be necessary to integrate the theory of sensing with models of collective migration to predict quantitatively what groups of migratory cells can and cannot do. Finally, controlled experiments with metastatic cells are required to validate these predictions, and to assess the viability of alternative therapies that target specific cell functions in order to combat metastatic invasion. Our hope is that the integrated, physics-based perspective presented herein will help generate innovative solutions to the pervasive problem of metastatic disease.

\section{Acknowledgments}

This work was supported by the Ralph W.\ and Grace M.\ Showalter Research Trust and the Purdue Research Foundation.


\end{document}